\documentclass[conference]{IEEEtran}
\IEEEoverridecommandlockouts
\usepackage{cite}
\usepackage{float}
\usepackage{amsmath, amssymb, amsfonts}
\usepackage{graphicx}
\usepackage{textcomp}
\usepackage{xcolor}
\usepackage[utf8x]{inputenc}
\usepackage{lmodern}
\include{macros3}
\def\BibTeX{{\rm B\kern-.05em{\sc i\kern-.025em b}\kern-.08em
    T\kern-.1667em\lower.7ex\hbox{E}\kern-.125emX}}

\ifCLASSOPTIONcompsoc
  \usepackage[nocompress]{cite}
\else
  \usepackage{cite}
\fi
\ifCLASSINFOpdf
\else
\fi

\DeclareRobustCommand*{\IEEEauthorrefmark}[1]{%
  \raisebox{0pt}[0pt][0pt]{\textsuperscript{\footnotesize #1}}%
}

\begin{document}
%\title{A Low-cost Eye-tracker for Early Detection of Mild Cognitive Impairment}

\title{A Cost-Effective Eye-Tracker for Early Detection of Mild Cognitive Impairment}

%\makeatletter
%\newcommand{\newlineauthors}{%
%  \end{@IEEEauthorhalign}\hfill\mbox{}\par
%  \mbox{}\hfill\begin{@IEEEauthorhalign}
%}
%\makeatother

\author{
  \IEEEauthorblockN{%
    Danilo Greco\IEEEauthorrefmark{1} \hspace{.2cm}
    Francesco Masulli\IEEEauthorrefmark{1}\, \IEEEauthorrefmark{2} \hspace{.2cm}
    Stefano Rovetta\IEEEauthorrefmark{1}\, \IEEEauthorrefmark{2} \hspace{.2cm}
    Alberto Cabri\IEEEauthorrefmark{1}\, \IEEEauthorrefmark{2} \hspace{.2cm}
    Davide Daffonchio\IEEEauthorrefmark{1}
  }
  \IEEEauthorblockA{%
    \IEEEauthorrefmark{1}DIBRIS University of Genoa - Via Dodecaneso,  35 - 16146 Genoa (Italy)\\%\{a.b $|$ c.d\}@abc.com \\
    \IEEEauthorrefmark{2}Vega Research Laboratories s.r.l. - Via Ippolito d’Aste,  7/5 - 16121 Genoa (Italy)\\ francesco.masulli@unige.it
  }
}%end author

% make the title area

\maketitle

\begin{abstract}

This paper  presents a low-cost  eye-tracker aimed at carrying out tests based on a Visual Paired Comparison protocol for the early detection of Mild Cognitive Impairment. The proposed eye-tracking system is based on machine learning algorithms,  a standard webcam,  and two personal computers that constitute,  respectively,  the "Measurement Sub-System" performing the test on the patients
%to which display the observer to which the test will be administered will be positioned 
and the "Test Management Sub-System" used by medical staff for configuring the test protocol,  recording the patient data,  monitoring the test and storing the test results. The system also integrates an stress estimator %detector 
based on the measurement of heart rate variability obtained with photoplethysmography. 
% starting from the image of the observer's face.

\end{abstract}


\section{Introduction}

%Eye-tracking is the process of measuring either the point of gaze (where one is looking) or the motion of an eye relative to the head. An eye-tracker is a device for measuring eye positions and eye movement. 

%Eye-trackers are devices aimed at  measuring either the direction of the of gaze (where one is looking) or the eye movements relative to the 
%Nowdays,  they are wispeadly used in many fields,  ranging from neuroscience,  to marketing,  to product design,  and  to man-machine interaction. 

%There are a number of methods for measuring eye movement. The most popular variant uses video images from which the eye position is extracted. %Other methods use search coils or are based on the electrooculogram.

Eye-trackers are devices designed to measure the direction of the gaze (where one is looking) and/or both eye movements related to the head.
Nowadays,  they are widely used in many fields,  ranging from neuroscience,  marketing,  product design and human-machine interaction~\cite{Romano14}.%\cite{Holmqvist15, Romano14}.

There are several methods for measuring eye movements,  including search coils~\cite{Robinson63}%\cite{Robinson63, Tumanski07}
 and electrooculograms~\cite{Keren10},  but currently,  the most popular eye-trackers use video images from which the eye position is extracted~\cite{Schreiber04, Houben06}.

%Several studies have linked cognitive impairments such as Alzheimer’s Disease (AD) with eye-related disorders and others indirectly related to the gaze~\cite{Javaid16}.
Some cognitive impairments,  such as Alzheimer’s Disease (AD),  are linked with eye movement-related disorders,  while others indirectly relate to the gaze dynamics~\cite{Javaid16}. Several investigations have been conducted in the last few years aimed at applying  eye-tracking technology to the detection of AD.

The main objective of medical research on AD is to diagnostic the disease as soon as
possible to maximize the effectiveness of therapies~\cite{LAGUN2011196}. %~\cite{LAGUN2011196, GAUTHIER20061262}. 
It is even possible to identify this disorder  before symptoms appear by detecting Mild Cognitive Impairment
(MCI).  Patients with symptoms of MCI have a very high risk of developing AD, 
with a conversion rate of 6\% to 25\% per year~\cite{Petersen99}. 
Although many patients with MCI tend
to become stable after some time,  more than half of them regress into dementia within
five years,  which outlines MCI as an excellent alarm bell that anticipates AD development.

Even if MCI does not markedly impair the daily activities of a person, 
some symptoms consist in  difficulties in carrying out complex operations that previously took
place without problems (such as preparing a meal)~\cite{GAUTHIER20061262}. 
It is therefore extremely important to identify symptoms of MCI from its the early stages.
Diagnostic biomarkers able to identify the pathology associated with this disorder include the detection of altered levels of
tau and amyloid in cerebrospinal fluid,  the use of structural Magnetic Resonance Imaging (MRI) to identify disease-specific patterns
of regional atrophy and MRI T1$\rho$ to detect disease-related macro-molecular protein aggregation,  and
the direct imaging of amyloid deposits using positron emission tomography and single photon
emission computerized tomography~\cite{Clark08}. 

As  it is not possible to foreseen a wide diffusion of all those 
%the listed above 
diagnostic tests to the entire adult population,   given their high costs and  degree of invasiveness,  eye-tracking-based tests are very promising in order to obtain reliable,  low-cost and non-invasive biomarkers for the early detection of MCI~\cite{zola2013}. In fact,  there are at least two types of tests based on eye-tracking that allow us to obtain bio-markers for MCI:
 
\begin{itemize} 
\item \emph{Saccadic Eye Movements (SEM) analysis}\\
Saccades are rapid eye movements of the eye followed by fixations,  that is,  the time in which the look is still to send information to
the brain and proceed to the next saccade. These movements are necessary and customarily used to perform actions such as reading.
Patients with AD presents irregularities of the saccades%~\cite{Leigh04} 
and  show difficulties reading the texts. These disorders are already present in a very early stage of AD%~\cite{Fernandez13} 
and in MCI before AD regression~\cite{Crawford05,  Opwonya22}.
%\begin{figure}[b!]
%\centering
%\fbox{\includegraphics[width=\columnwidth,  height=200px,  keepaspectratio]{images/test_drawing.png}}
%\caption{Example of three possible architecture configurations,  respectively a PC connected to another laptop,  an all-in-one pc or a micro-pc with a screen and a webcam.}observer
%\label{fig:test_drawing}
%\end{figure}

\item  \emph{Visual Paired Comparison (VPC)}\\
This type of test~\cite{LAGUN2011196} relies on memory impairment caused by MCI. The VPC tests may consist into two distinct
phases: the familiarization phase and the test phase. 
Both test steps display pairs of images for a assigned time on a  screen,  using the eye-tracker to monitor the observer's fixations. As an example,  
the pictures are presented in pairs with %offer an 
identical stimulus during the familiarization phase,  while 
during the test phase,  the pairs of images to be displayed are created by placing on one side of the screen  a known image already seen in the familiarization phase and on the other a new image. 
Healthy people usually observe the new one 70\% of the 
stimulus duration. Patients with MCI,  on the other hand,  do not give preference to either of the two stimuli,  reinforcing the hypothesis that they do not remember the images of the familiarization phase. These data highlight the short-term memory damage caused by MCI~\cite{Crutcher09}.
VPC tests are versatile and simple and they  have  been used with rodents,  %~\cite{Clark00}, 
 primates,  %~\cite{Zola00},  
 children, %~\cite{Fagan90}
  and adults~\cite{Manns00}.~\cite{Zola00} showed that the VPC tests are  able to diagnose brain damage caused by MCI not evidenced by MRI.
\end{itemize}

We note that a test for the early diagnosis of Mild Cognitive Impairment based on Visual Paired Comparison requires the measurement of the duration of the fixations on the images presented on the left and on the right side of a screen. As fixations in eye movements are events lasting at least 100 ms~\cite{Nuthmann17},  it is therefore possible to use for eye-tracking a standard webcam  with a frame rate of 30 fps.

\begin{figure}[b]
\centering
\includegraphics[width=\columnwidth,  height=200px,  keepaspectratio]{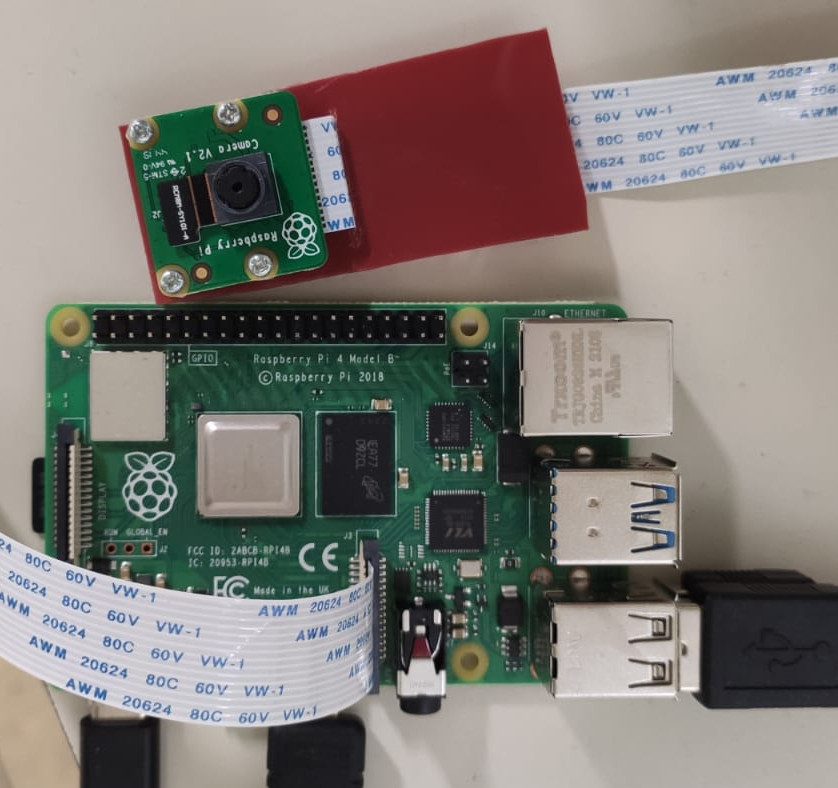}
\caption{Raspberry Pi 4 and Raspberry Pi Camera Module 2.}
\label{fig:raspberry}
\end{figure}

\begin{figure}[b]
\centering
\fbox{\includegraphics[width=\columnwidth,  height=114px,  keepaspectratio]{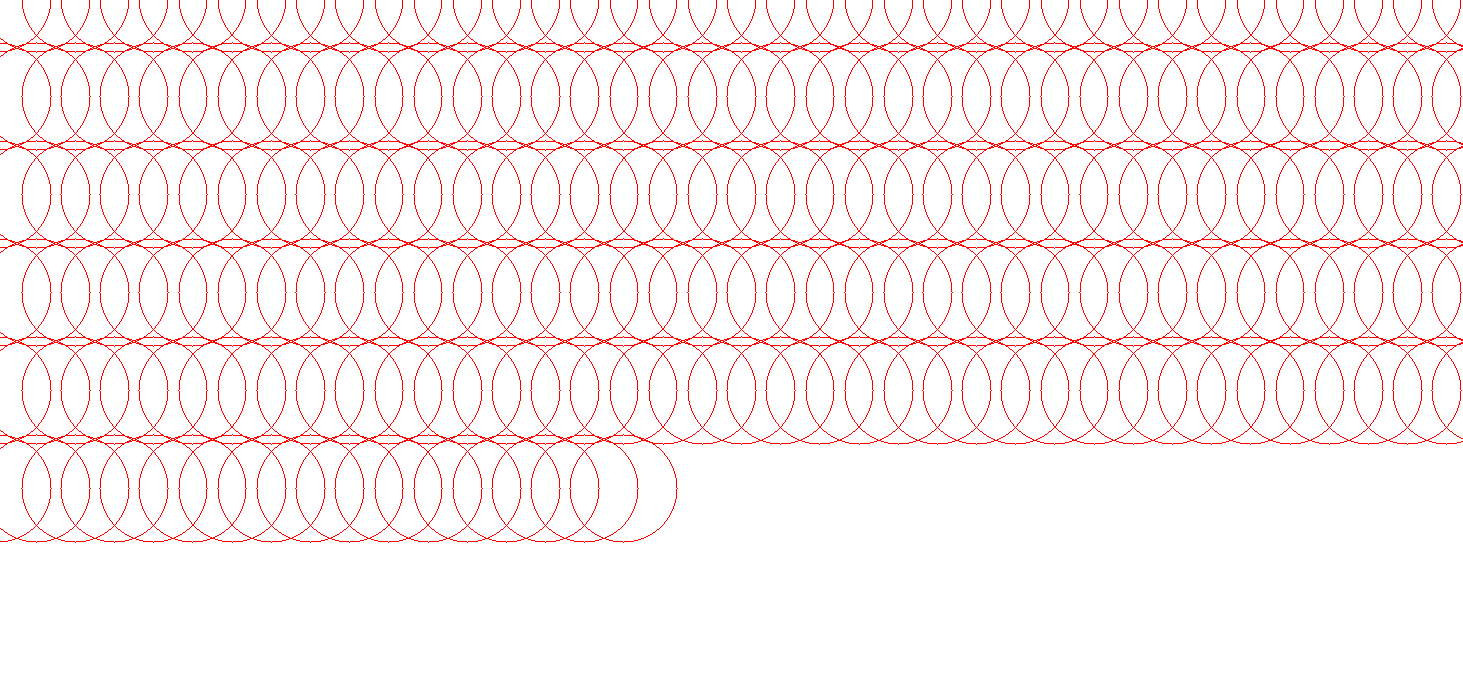}}
\caption{Circle following a predefined pattern,  used for calibration.}
\label{fig:training}
\end{figure}

\begin{figure}[b]
\centering
\includegraphics[width=\columnwidth,  height=150px,  keepaspectratio]{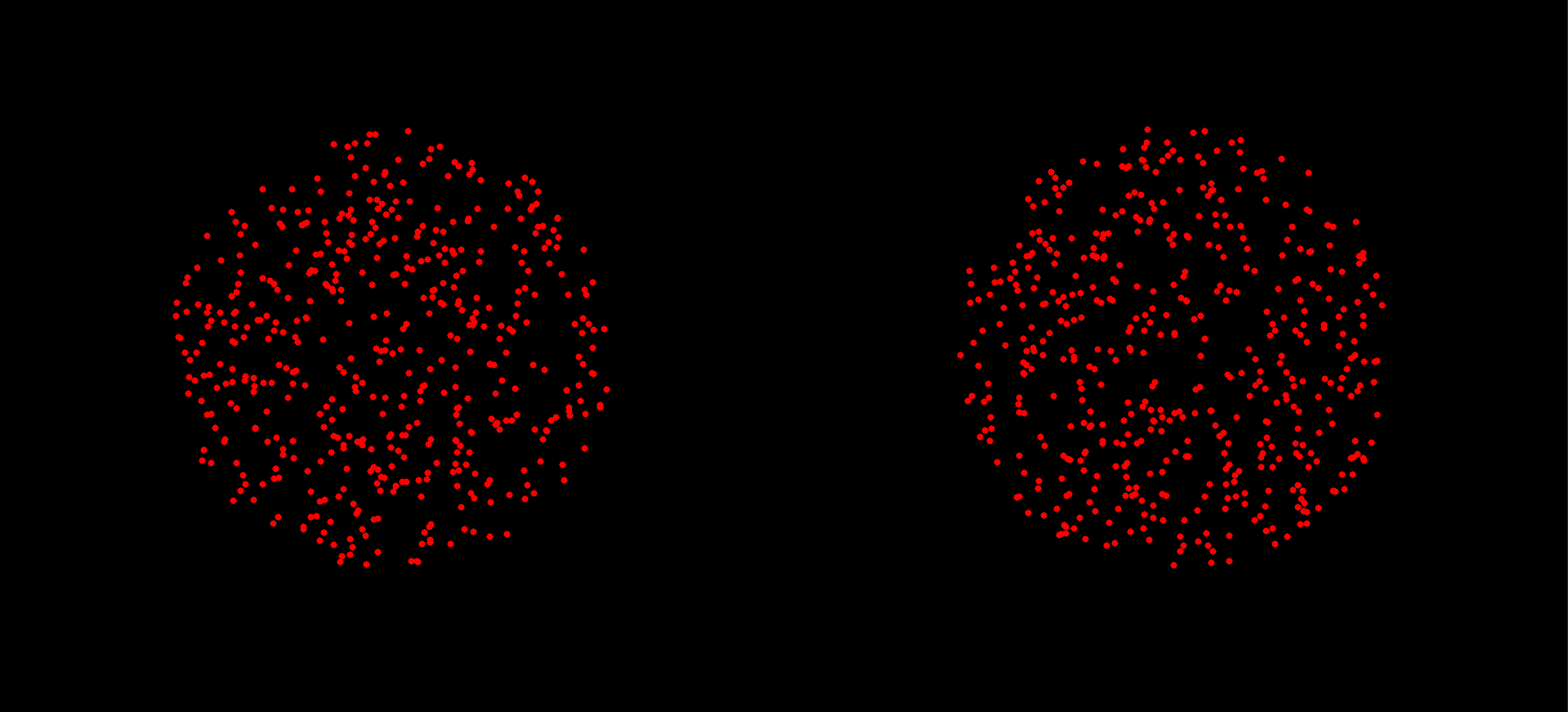}
\caption{Click-points pattern used to measure the system's accuracy.}
\label{fig:testing_accuracy}
\end{figure}

\begin{figure}[t]
\centering
\includegraphics[width=\columnwidth,  height=150px]{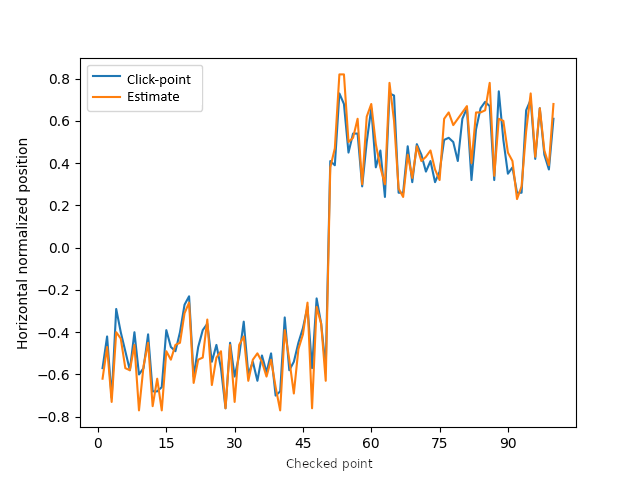}
\caption{
%Graphic comparison of the tests carried out with the two last versions of the neural network (with the final version on the bottom). 
Accuracy evaluation: comparison of the horizontal position of the click-point vs the neural network estimate.}
\label{fig:test_plot}
\end{figure}

\begin{figure}[b]
\centering
\includegraphics[width=\columnwidth,  height=240px,  keepaspectratio]{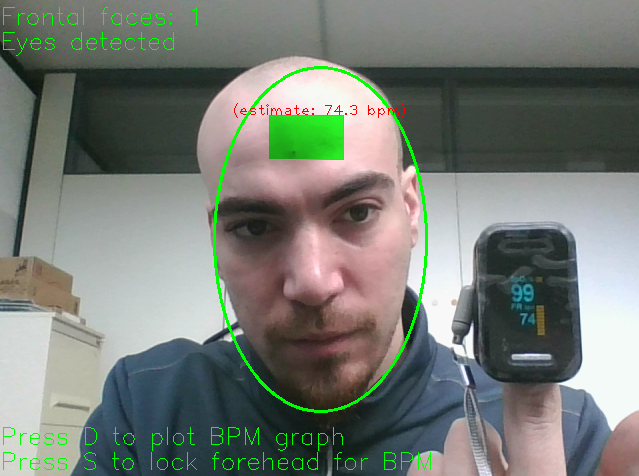}
\caption{Calibration phase: the ellipse turns green if there is a face inside and the eyes are detected. BPM is also estimated. This figure compares also the optical heart rate monitor of our system with a commercial pulse oximeter device.}
\label{fig:calibration}
\end{figure}

\begin{figure}[t!]
\centering
\includegraphics[width=\columnwidth]{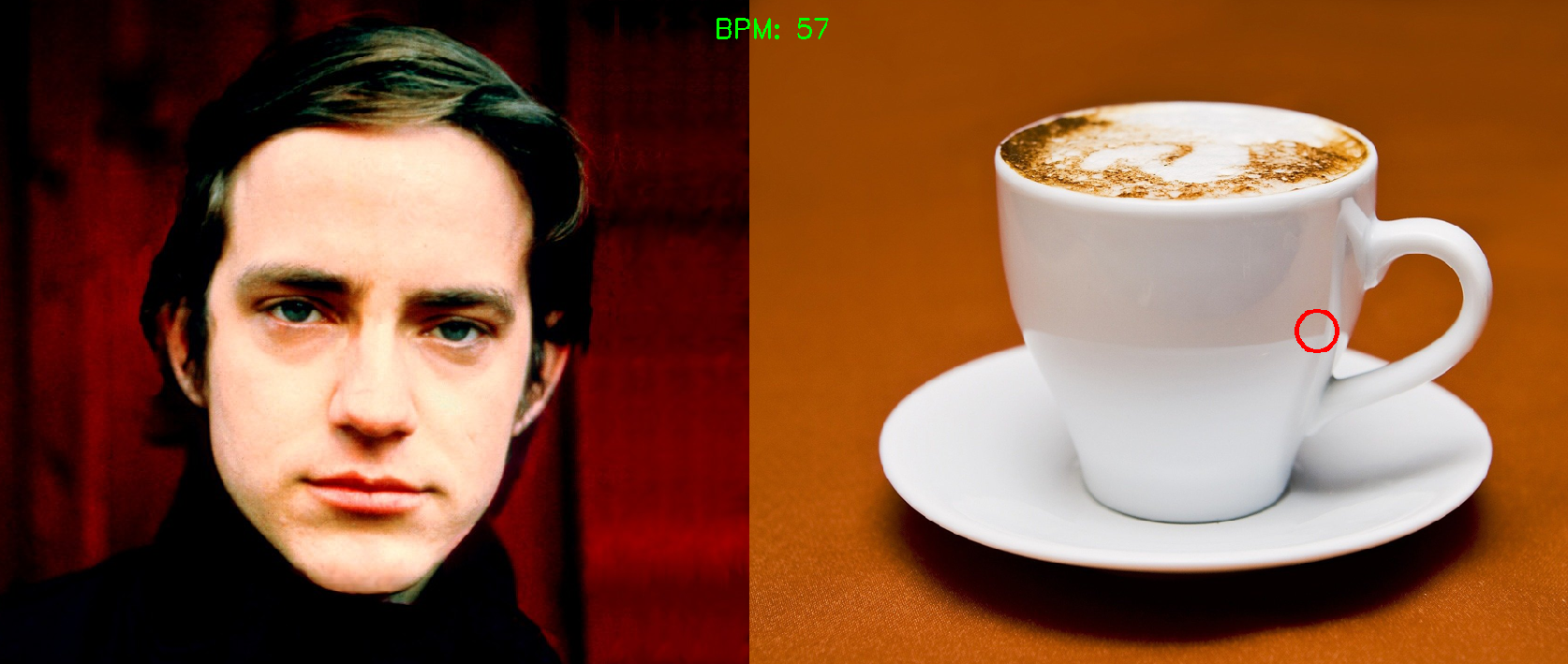}
\caption{Test Management Sub-System view: The medical staff can see the images visualized on the screen of the patient,  augmented with position of the gaze marked with a red circle and the heart rate  (in this case,  the patient is watching the middle of the cup).}
\label{fig:test}
\end{figure}

Starting from this consideration,  we designed an eye-tracking system based on two computers and a standard webcam that implements a Visual Paired Comparison for the early diagnosis of the Mild Cognitive Impairment. The two computers constitute,  respectively,  the "Measurement Sub-System" and the "Test Management Sub-System". 
The "Measurement Sub-System" consists of a desktop computer with a display of at least 24" with at least a I3 CPU and a webcam (or,  better,  an all-in-one computer with the same characteristics),  or,  also,  a single-board computer such as a Raspberry (see Fig. ~\ref{fig:raspberry}) attached to a webcam and a display of at least 24". The observer to whom the test is administered is positioned in front of the display. 
 The "Test Management Sub-System" is a desktop or a laptop with at least a I3 CPU and is used by medical staff for configuring the test protocol,  recording the patient data,  monitoring the test and storing the test results. 

The rest of this paper is organized as follows: Sec. II  presents the technical aspects of the  system  we designed and the evaluation of its performance; Sec. III contains an example of protocol of a test for the early diagnosis of Mild Cognitive Impairment based on Visual Paired Comparison; Discussion and Conclusions are depicted in Sec. IV.

\section{System Implementation}

%%%%%%%%%%%%%%%%%%%%%%%%%%%%%%%%%%%%%%%%%%%%%%

\subsection{System Design}
Our approach to the eye-tracking design is based on the usage of video images from a webcam from which the gaze point on the screen is estimated~\cite{Schreiber04, Houben06}.

The necessary steps for gaze point measurement are the detection the position of the observer's face inside the images acquired by the webcam and then the position of the eyes into them and the estimation of the observed point on the screen.

The software was developed using the Python language~\cite{van2007python}%, culjak2012brief}  
under the Linux operating system. It is based
on two %Haar feature-based cascade classifiers
classifiers exploiting the  Viola–Jones object detection framework~\cite{viola2004robust, lienhart2002extended},  the former used to detect the position of the observer's face and the latter used to detect the position of the eyes and on a dense neural network used for calibration~\cite{zemblys2018using}.
%Our solution and approach (see Figs.~\ref{fig:raspberry}) uses a low-cost commercial webcam leveraging on the neural networks to compensate for the limited performances of the cheapest hardware used. 
%\begin{figure}[b!]
%\centering
%\includegraphics[width=\columnwidth]{images/autism_images.png}
%\caption{An example of a pair of images used in a possible test for autism (a face and a cup of coffee).}
%\label{fig:autism_images}
%\end{figure}
The Viola–Jones object detection framework is a robust and real-time algorithm   primarily motivated by the problem of face detection that is  composed by four stages:
%with is an object detection framework which was proposed in 2001 by Paul Viola and Michael Jones.[1][2] Although it can be-trained to detect a variety of object classes,  it was motivated primarily by the problem of face detection.
\begin{enumerate}
\item Haar Feature Selection,  using  Haar basis functions~\cite{Haar1910};
\item Creating an Integral Image~\cite{Crow84, Lewis95}; 
\item Adaboost Training~\cite{Schapire99};
\item Cascading Classifiers. 
\end{enumerate}

%\begin{figure}[b!]
%\centering
%\fbox{\includegraphics[width=\columnwidth,  height=150px,  keepaspectratio]{images/neural_network.png}}
%\caption{Graphic representation of a neural network. The input (in our case the image of the eye) passes through various internal layers of neurons,  multiplied by weights with the addition of a bias and normalized with an activation function. The output is the predicted direction of the gaze.}
%\label{fig:neural_network}
%\end{figure}

%\begin{figure}[b!]
%\centering
%\includegraphics[width=250px,  height=150px ]{images/rt_test.png}
%\caption{Example of instant feedback testing. The red circle will move following the observer's gaze. In this case,  the observer is looking left. Notice that the webcam is mirrored.}
%\label{fig:rt_test}
%\end{figure}

The Haar feature-based cascade classifiers~\cite{viola2004robust, lienhart2002extended} are provided by a function of the OpenCV library that is a large open-source library for computer vision that we use also for images and webcam frames  manipulation and for the  implementation of the graphical user interfaces~\cite{culjak2012brief},   together with the PyQT library~\cite{pyqt_docu12},  while the neural network is implemented using Tensorflow~\cite{tensorflow2015-whitepaper}. %(in alternative to PyQT that is Qt is not natively compatible with the Raspberry architecture,  and displaying images of any type is much more efficient using OpenCV.

%The neural network is sequential and dense,  and it is structured as follows. It has an input layer of 1024 neurons,  two hidden layers with 100 and 10 neurons,  and an output layer of 1 neuron. The shape of the input data defines the input layer quantity of neurons,  that is,  a 32x32 picture containing the eye. The output is a scalar value between -1 and 1,  so we need just a single neuron. The number of neurons in the middle layers is arbitrary and follows generic rules. The neural network is dense because it fits more effectively with our problem. While convolutional models are better with a classification algorithm,  dense models are better with scalar values. 

%We also use a standard SVG optimizer.

\subsection{Calibration procedure and assesment}

Before the actual test it is  necessary to carry out for each observer a calibration procedure where the geometric characteristics of the observer's eyes are (implicitly) estimated as the basis for a fully-customized and accurate gaze point prediction.

During calibration,   a blank screen is presented to the observer  with a circle that moves following a predefined pattern (Fig.~\ref{fig:training}). The observer is requested to keep his face still,  with eyes opened and visible,  the face frontal and centered with respect to the camera and to pursuit the circle with the eyes; moreover,  only a single face must appear in the frame. If one of these conditions are not met,  the program gives an alarm and goes into stand-by,  waiting for the observer to position himself correctly again.

The eye images and the effective positions of the stimuli on the screen are collected in a data set that is used for training a dense neural network~~\cite{Goodfellow16} aimed at creating a (in general non-linear) map that will allow to estimate the positions of the gaze points on the screen during the test.

%To this end,  the observer is requested to pursuit with eyes a sequence of stimuli drawn on the screen. The measured positions are compared with the effective positions of the stimuli on the screen,  in order to create a (generally non-linear) map that  allows to obtain the correct positions of the points observed on the screen.

The architecture of the  dense neural network comprises an input layer of 1024 neurons acting on the sub-image 32x32  containing the eye,  two hidden layers of 500 and 100 ReLU units and an output layer composed by one sigmoidal neuron giving the estimate horizontal position of the observed point on the screen.  As the training is aimed at interpolation,   we  bring the optimization procedure into over-fitting (using a large number of epochs) and then obtain a precise localization of the observed point on the screen.

To quantitatively measure the accuracy of the neural network,  a test procedure has been developed in which the observer is required to click on points that appear one-by-one in random positions on the right or left side of the screen. Based on the answers given by the observer,  the program gives an evaluation of the accuracy on the right / left side discrimination (see Fig.~\ref{fig:testing_accuracy}).

%We implemented a Multi-observers Calibration (MSC) procedure on 21 different observers (10 males,  11 females) and then performed the test of accuracy described in Sect.II a) on seven additional observers (4 males,  3 females) taking 30 completely random points on the screen. The result showed a 15\% average absolute error (about 5 cm on the screen of 15 inches used) and the ability to distinguish the right from the left 90\% of the time.

%Later,  we improved some features of the program in the following way. First,  the-training system took a higher amount of data in less time (using a circle that moves automatically in a given pattern as described above rather than having the observer click points on the screen as in the old version of the software). The accuracy-test system selected points on the screen in a smarter way (as described in the Neural Network Training and Testing sub-section) and no longer randomly. 

%Better results where obtained by using a Single-observer Calibration (SSC) procedure on a single observer.

The test of accuracy on 100 points  (50 on the left and 50 on the right of the screen) shows an average absolute error of 3.00\% (1.6 cm on a 24-inch screen),  with an ability to distinguish the right from the left of 100\% (see Fig.~\ref{fig:test_plot}).

\subsection {Heart Rate Variability Monitoring}

%%%%%%%%%%%%%%%%%%%%%%%%%%%%%%%%%%%%%%%%%%%%%%%%%%%%%%%%%%%%%%%%%%%%%%%%%%%%%%%%%%%%%%%

It's worth noting that   it is possible to exploit the proposed architecture to add additional functionalities useful for monitoring  the mental state of the patient during the MCI test,  such as facial expression detection~\cite{Bartlett03},  and stress monitoring through the measurement of heart rate~\cite{Tiba2013ImageBasedAP} or heart  beats per minute (BPM).

%s the collection of other measures related to the psychophysical state of the observer,  such as the heart rate and facial expression. 

%The final goal is to distinguish through this application whether the observer is looking left or right on the screen as a potential tool to perform tests for early detection of symptoms related to Alzheimer's disease (AD) and autism. 

In our system,    a region of interest (ROI) $\Omega_{\mathrm{fROI}}$ located in the forehead is selected
to estimate the heart rate~\cite{Tiba2013ImageBasedAP}. This is done  by measuring in the sRGB color space the variation of the average optical intensity in the region in the e green channel that contains the most pulse information~\cite{Verkruysse:08}
\begin{equation}
C_{\mathrm{PPG}}^{i}(t)=\frac{\sum_{x,  y \in \Omega_{\mathrm{ROI}}} C^{i}(x,  y,  t)}{\left|\Omega_{\mathrm{fROI}}\right|},  \; \; i \in\{\mathrm{R},  \mathrm{G},  \mathrm{B}\}
\end{equation}
where $C^{i}(x,  y,  t)$ is the pixel value at the coordinates $(x,  y)$ in the $i$ channel at the time $t$.%~\cite{ryu2021research}.
%and $|\Omega_{\mathrm{fROI}}|$ represents the area of the facial region of interest (ROI) 

This feature works thanks to the photoplethysmography (PPG) technique, %~\cite{allen2007photoplethysmography},  
which can capture optical absorption characteristics of (oxy-)hemoglobin~\cite{Verkruysse:08}. In short,  when the heart pumps,  the average optical intensity in the ROI changes. This change %is invisible to the naked eye,  though it 
can be caught  by a standard webcam.%~\cite{zhang2021noncontact}.
%. In the green channel of the image,  this change is even more obvious. 
By processing this information,  we can estimate the patient's heart rate and then the heart rate variability (HRV) that is the fluctuation of the length of heart beat intervals.  HRV is closely related to emotional arousal and stress~\cite{Kim18}.%s~\cite{taelman2009influence, Kim18}.

%%%%%%%%%%%%%%%%%%%%%%%%%%%%%%%%%%%%%%%%%%%%%%%%%%%%%%%%%%%%%%%%%%%%
 In Fig.~\ref{fig:calibration} we show the visual interface presented by the "Test Management Sub-System" during the calibration procedure. We noticed the face of the observer enclosed in a green ellipse,  as he is well positioned. The BPM are also visualized; for sake of clarity we have added a pulse oximeter to check and compare the BPM evolution.

\subsection{Single-Board Computer Version}

Using a single-board computer,  such as the Raspberry PI4~\cite{raspberry} %(see Fig.~\ref{fig:raspberry}) 
for implementing the ”Measurement Sub-System”,  we encountered significant limitations in the processing speed. To overcome these limitations,  we included in our software the possibility that the ”Measurement Sub-System” locally records the frames from the camera while the observer watches the pairs of images. When the test is over,  the ”Measurement Sub-System”  transfers the complete data gathered by the camera to the ”Test Management Sub-System”,   which calculates and save in a database all the predictions on each frame and the corresponding heart rate.

%The "Patient" will locally record the frames from the camera while watching the pairs of images. Concurrently,  the "Doctor" will see the same pair of images that the "Patient" is watching and an estimation of the gaze direction of the "Patient, " represented by a red circle (see Fig.~\ref{fig:test}). 
 
% When the test is over,  the "Patient" will automatically transfer the complete data gathered by the camera to the "Doctor, " which will calculate and save all the predictions on each frame and the corresponding heart rate in a database.
 
%  We use this specific architecture to avoid complex and computationally expensive operations on the "Patient" machine,  to guarantee a smooth experience during the test even with very minimal hardware.

\subsection{Sub-systems synchrony}
%The architecture is composed of two different computers connected to the same network,  a more powerful machine used by the supervisor of the test (that we will call "Doctor"),  and another machine used by the "Patient". 

As already stated,  the proposed eye-tracking system is composed by the  ”Measurement Sub-System” aimed to interact with the patient and the ”Test Management Sub-System” used by medical staff for perform the calibration,  configuring the test protocol,  recording the patient data,  monitoring the test and storing the test results. 

The two subsystems are connected to each other in a sub-net created via a router or an access point and the "Test Management Sub-System" can also be connected to the network of the hospital.

During the test,   the same pair of images visualized on the screen of the  "Measurement Sub-System" will appear in synchrony on the screen of the "Test Management Sub-System",  augmented with the position of the gaze marked with a red circle (see Fig.~\ref{fig:test}) and the heart rate.

% The first one gives instant feedback,  showing in real-time the direction of the observer's gaze with a red circle (see Fig.~\ref{fig:rt_test}). This function gives an approximate visual idea of the proper functioning of the neural network. 

% The "Doctor" can start the procedure when everything is ready. At this point,  the software will show the next pair of images to the "Patient" in a predefined pattern decided by the "Doctor" (basically the order of the pair of images and the amount of time of watching).

%Our software is optimized to work with the "Patient" being a cheap machine (like a Raspberry~\cite{raspberry}Fig.~\ref{fig:raspberry}). - on today - see  During the test,  the "Doctor" has total control of the procedure and continuously receives visual feedback from the "Patient, " such as the camera's streaming. 

\begin{figure}[t]
\centering
\includegraphics[width=\columnwidth]{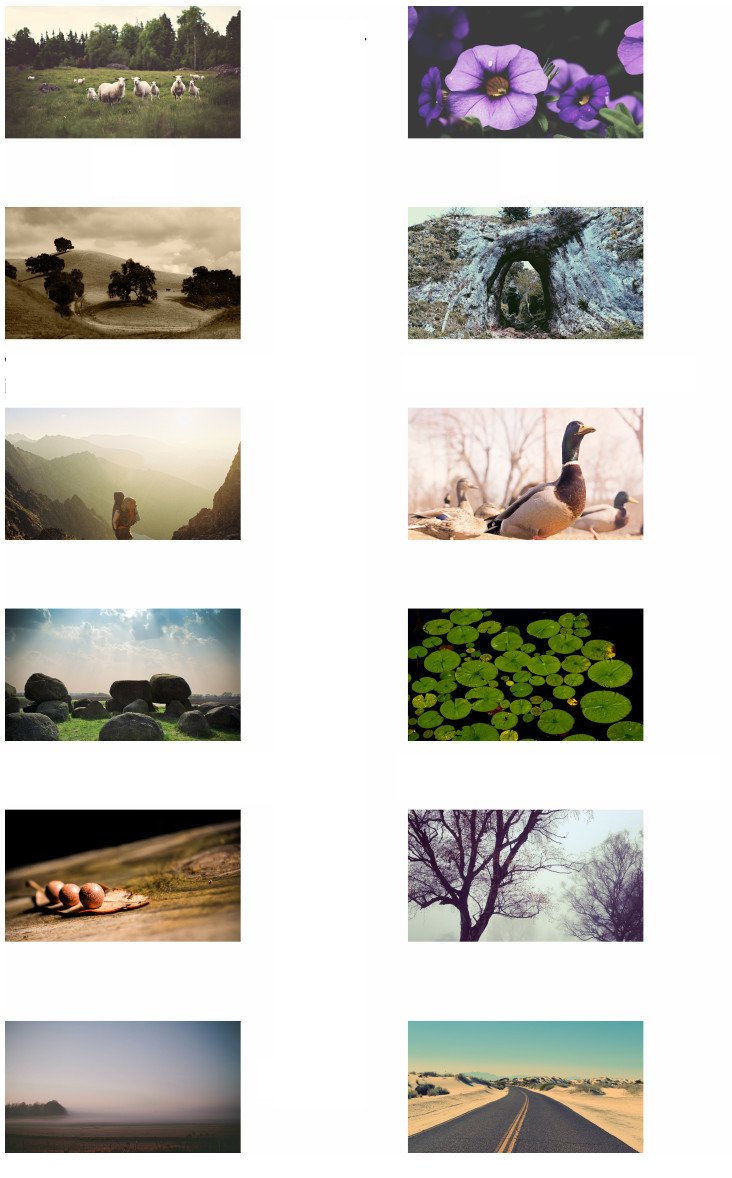}
\caption{MCI Visual Paired Comparison  test: Images used in the familiarization phase.}
\label{f-images}
\end{figure}

\begin{figure}[t]
\centering
\includegraphics[width=\columnwidth]{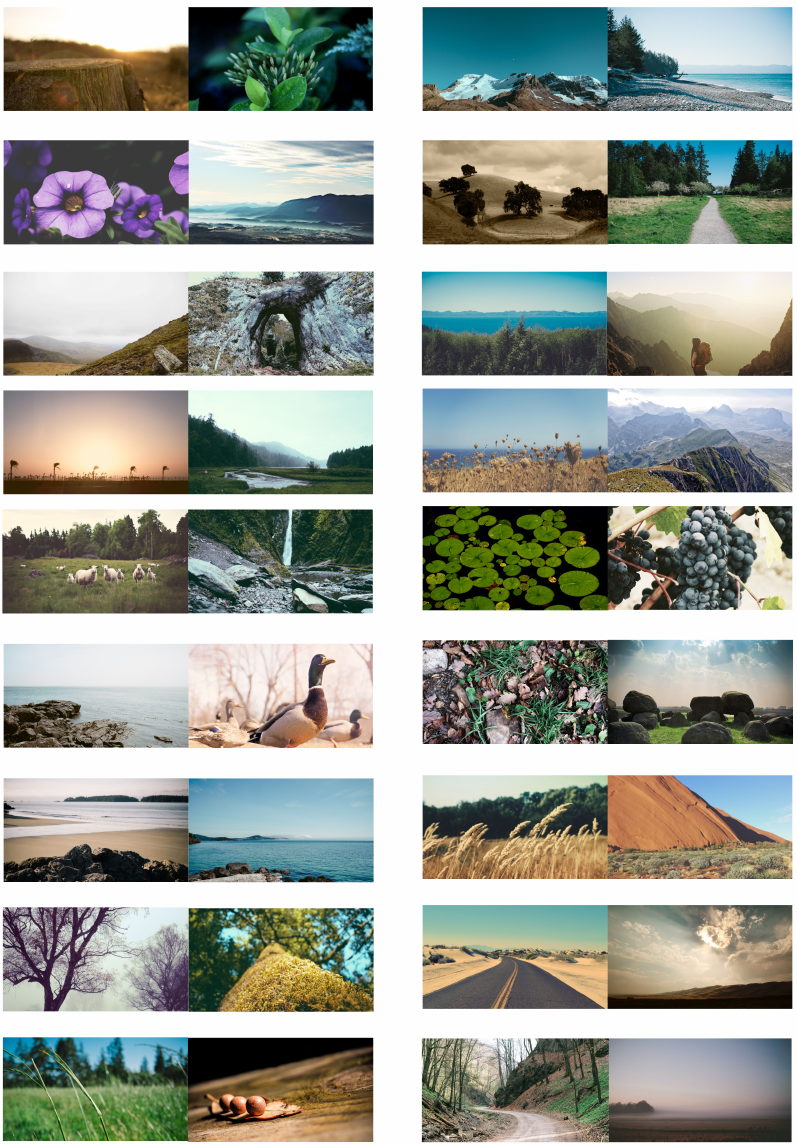}
\caption{MCI Visual Paired Comparison  test: Examples of pair of images displayed in the test phase.}
\label{t-images}
\end{figure}

\section{Example of Test Protocol}

%%%%%%%%%%%%%%%%%%%%%%%%%%%%%%%%%%%%%%%%%%%%%%%%%%%%

In collaboration with the physicians of the Neurological Clinic of the University of Genoa,  we are defining the details of a protocol for the implementation of a test for the early diagnosis of Mild Cognitive Impairment based on Visual Paired Comparison,  with the aim of testing it in clinics with geriatric patients.

A possible protocol employs 36 images,  of which 12 are presented in the familiarization phase,  and 24 in the testing phase only

During the familiarization phase (total duration 3 minutes) the images are shown on the display for 15 seconds each,  accompanying it with a specific question aimed at reinforcing its memorization,  e.g.,   ”What kind of animal is shown?” (see Fig.~\ref{f-images}).
After the familiarization phase,  the test phase starts (see Fig.~\ref{t-images}). In this phase (total duration 4 minutes) we   show in random sequence:
\begin {itemize}
\item six pairs of new images;
\item six pairs with a new image on the left and a known image from the familiarization phase on the right;
\item six pairs with a new image on the right and a known image from the familiarization phase on the left.
\end {itemize}
As already discussed,  we expect that healthy people will observe the new stimuli about 70 \% of the time,  while observers with MCI will not give preference to either of the two stimuli~\cite{Crutcher09}.

\section{Discussion and Conclusions }

In this paper we have presented a system for eye-tracking aimed at carrying out tests based on a Visual Paired Comparison protocol for the early detection of Mild Cognitive Impairment. The proposed system uses webcams,  standard personal computers that may already be available at hospital structures (or even low cost-single board computers) and exploits some efficient machine learning algorithms.

Even if the proposed system is not the first affordable eye-tracker that uses artificial intelligence algorithm,  standard webcams and personal computers (see,  e.g., ~\cite{Corno02, zemblys2018using}),  it is characterized by its specialization and validation for tests based on Visual Paired Comparison protocol and for  its integration with stress measurement based on heart rate variability. Given the low-cost of the proposed eye-tracking system,   once the measurement protocol has been defined and its clinical trials concluded,  a wide diffusion of the test is expected,  with great positive repercussions on public health,  given that the Alzheimer's disease is one of the chronic diseases that affects a large segment of the elderly population and is spreading more and more also in low-income countries,  with the rise in life expectancy.

\bibliographystyle{unsrt}
\bibliography{masulli-melecon2022.bib}

\end{document}